\documentclass[aps,prb,reprint]{revtex4-1}

\usepackage{latexsym,amssymb,float}
\usepackage{setspace}
\usepackage{graphicx}
\usepackage{epsfig}
\usepackage{amsmath}
\usepackage{bm}
\usepackage{appendix}
\usepackage{verbatim}

\begin{document}

\title{Scanning Tunneling Shot Noise Spectroscopy in Kondo systems}

\author{Sagen Cocklin and  Dirk K. Morr}

\affiliation{Department of Physics, University of Illinois at Chicago, Chicago,
IL 60607}
\date{\today}

\begin{abstract}
Using a large-N theory in combination with the Keldysh non-equilibrium Greens function formalism, we investigate the current, differential conductance, zero-frequency shot-noise and Fano factor as measured by scanning tunneling shot noise spectroscopy (STSNS) using a scanning tunneling microscope (STM) near single Kondo impurities and in Kondo lattices. We show that the Fano factor $F$ exhibits a characteristic bias dependence arising from Kondo screening that is similar to the Kondo resonance observed in the differential conductance. Moreover, the lineshape of $F$ is strongly dependent on the ratio of the tunneling amplitudes for electron tunneling from the STM tip into the conduction band and electronic levels of the magnetic adatoms. We demonstrate that the Fano factor can be enhanced or suppressed due to interference effects and as such, is not only a sensitive probe for the correlation effects arising from Kondo screening, but also for quantum interference between tunneling electrons. We identify a correlation between the form of the differential conductance and the Fano factor that could be tested in future STSNS experiments.

\end{abstract}

\date{\today}

\maketitle

\section{Introduction}

The Kondo screening of a magnetic impurity by conduction electrons is one of the most fascinating phenomena in condensed matter physics \cite{Kon64}. Its local spectroscopic signature, the Kondo resonance, has been well studied using scanning tunneling spectroscopy (STS) experiments \cite{Mad98,Li98,Man00,Mad01,Kno02}. By measuring the local differential conductance near magnetic adatoms such as a Co atom located on metallic Cu(111) or Au(111) surfaces, it was observed that the Kondo resonance exhibits a lineshape (i.e., a bias dependence) with a characteristic asymmetry that can be well described phenomenologically using the Fano formula \cite{Fano61}. A microscopic derivation of the Fano formula \cite{Fano61,Fig10,Morr17} has shown that the asymmetry of the Kondo resonance arises not only from the particle-hole asymmetry of the underlying conduction band, but also from quantum interference between electrons tunneling from the STM tip either into the conduction band, or the electronic levels of the magnetic adatoms \cite{Mal09} as schematically shown in Fig.~\ref{fig:STM}.

The recent progress \cite{Bur15,Mas18,Bas18a,Bas18b,Mas19} in the development of scanning tunneling shot noise spectroscopy (STSNS) \cite{Birk95,Kem07,Herz13} using a scanning tunneling microscope (STM) in which an STM tip is used to simultaneously measure the $IV$-curves as well as the zero-frequency shot noise -- the current-current correlation function -- has raised the question of whether such a characteristic signature as the Kondo resonance can also be found in the bias dependence of the shot-noise, or of the Fano factor, defined as
\begin{equation}
F=\frac{S(\omega=0)}{2e|I|}
\label{eq:Fano}
\end{equation}
where $I$ is the current flowing from the tip into the system, and $S(\omega=0)$ is the associated zero-frequency shot-noise. Indeed, recent STSNS experiments have found a strong suppression of the Fano factor from its Poissonian value of unity around magnetic and non-magnetic adatoms on a Au(111) surface \cite{Bur15}, and observed an enhanced current noise near defects in cuprate superconductors \cite{Bas18b,Mas19}. Furthermore, it was argued that the observation of shot-noise via STSNS could provide insight into the local spin susceptibility associated with unscreened magnetic adatoms \cite{Nus03}, and that measurements of conductance-conductance correlations using an STM tip could provide insight into the local spin structure of the Kondo screening cloud \cite{Pat09}.

Shot-noise and the Fano factor have been extensively studied in mesoscopic systems \cite{Bla00,Zhu03,Sot10}, in particular in the context of the Kondo effect on a quantum dot \cite{Meir02,Dong02,Lop03,Lop04,Wu05,Sela06,Zar08,Mora08,Yam11,Fer15} or in carbon nanotubes \cite{Del09}, in a set-up that is qualitatively different from that of STSNS experiments. While it was shown that the Fano factor in quantum dot systems is suppressed by Kondo correlations \cite{Dong02}, it was also predicted \cite{Sela06} that in the unitary limit, the Fano factor associated solely with the backscattered current can exceed the Poissonian limit of unity. An enhancement in this modified Fano factor was subsequently observed experimentally \cite{Zar08,Yam11}.

In this article, we will investigate the relation between the bias dependence of the current, the differential conductance, the zero-frequency shot-noise, and the Fano factor around magnetic adatoms located on metallic surfaces, exhibiting a Kondo effect, as well as in Kondo lattices, as observed by scanning tunneling shot noise spectroscopy. We will show that the Fano factor exhibits a characteristic lineshape that reflects not only the strong correlations arising from Kondo screening, but also quantum interference effects due to multiple tunneling paths. This characteristic lineshape of $F$ is not unlike the Kondo resonance observed in the differential conductance, and presents an additional test for our understanding of the Kondo effect.

The rest of the paper is organized as follows. In Sec. \ref{sec:theory} we present our theoretical model and derive the form of the current and shot-noise measured by an STM tip. This model was previously employed to successfully describe the Kondo resonance of a Co adatom located on a Au(111) surface. In Sec.~\ref{sec:SN_Kimp} we discuss our results for the shot noise around a single magnetic adatom, the relation between the differential conductance and the shot noise lineshape, and the effects of tunneling interference. In Sec.~\ref{sec:SN_KL} we discuss the form of the shot noise and Fano factor in Kondo lattice systems. Finally, in Sec.~\ref{sec:concl} we present our conclusions.

\section{Theoretical Model}
\label{sec:theory}

We begin by discussing the model for the current and shot noise measured by STSNS around a single Kondo impurity, and will subsequently extend it to the Kondo lattice. To study the properties of a single Kondo impurity, we employ the theoretical model of Ref.\cite{Fig10} which was used to successfully describe the lineshape of the Kondo resonance in the differential conductance, $dI/dV$, measured around a single magnetic Co adatom located on a metallic Au(111) surface \cite{Mad98}. Such a system is described by the Hamiltonian \cite{Kon64}
\begin{equation}
\hat{H} = - \sum_{\textbf{r},\textbf{r}',\sigma} t_{\textbf{rr}'}c^{\dag}_{\textbf{r},\sigma}c_{\textbf{r}',\sigma}  + J \textbf{S}_{\textbf{R}}^{K} \cdot \textbf{s}_{\textbf{R}}^{c}
\label{eq:H}
\end{equation}
where $c^\dagger_{{\bf r},\sigma}$  $(c_{{\bf r},\sigma})$ creates (annihilates) a conduction electron with spin $\sigma$ at site ${\bf r}$ on the Au(111) surface. Here, $t_{{\bf rr'}}=1.3$ eV is the fermionic hopping element between nearest-neighbor sites in the triangular Au(111) surface lattice, and $\mu= -7.34$ eV is its chemical potential. These parameters describe the dispersion of the experimentally observed  Au(111) surface state \cite{Sch09} that takes part in the Kondo screening of the Co adatom. Moreover, $J>0$ is the Kondo coupling, and ${\bf S}^{K}_{\bf R}$ and ${\bf s}^c_{\bf R}$ are the spin operators of the magnetic Co adatom and the conduction electron at site ${\bf R}$, respectively.

\begin{figure}
\includegraphics[width=5cm]{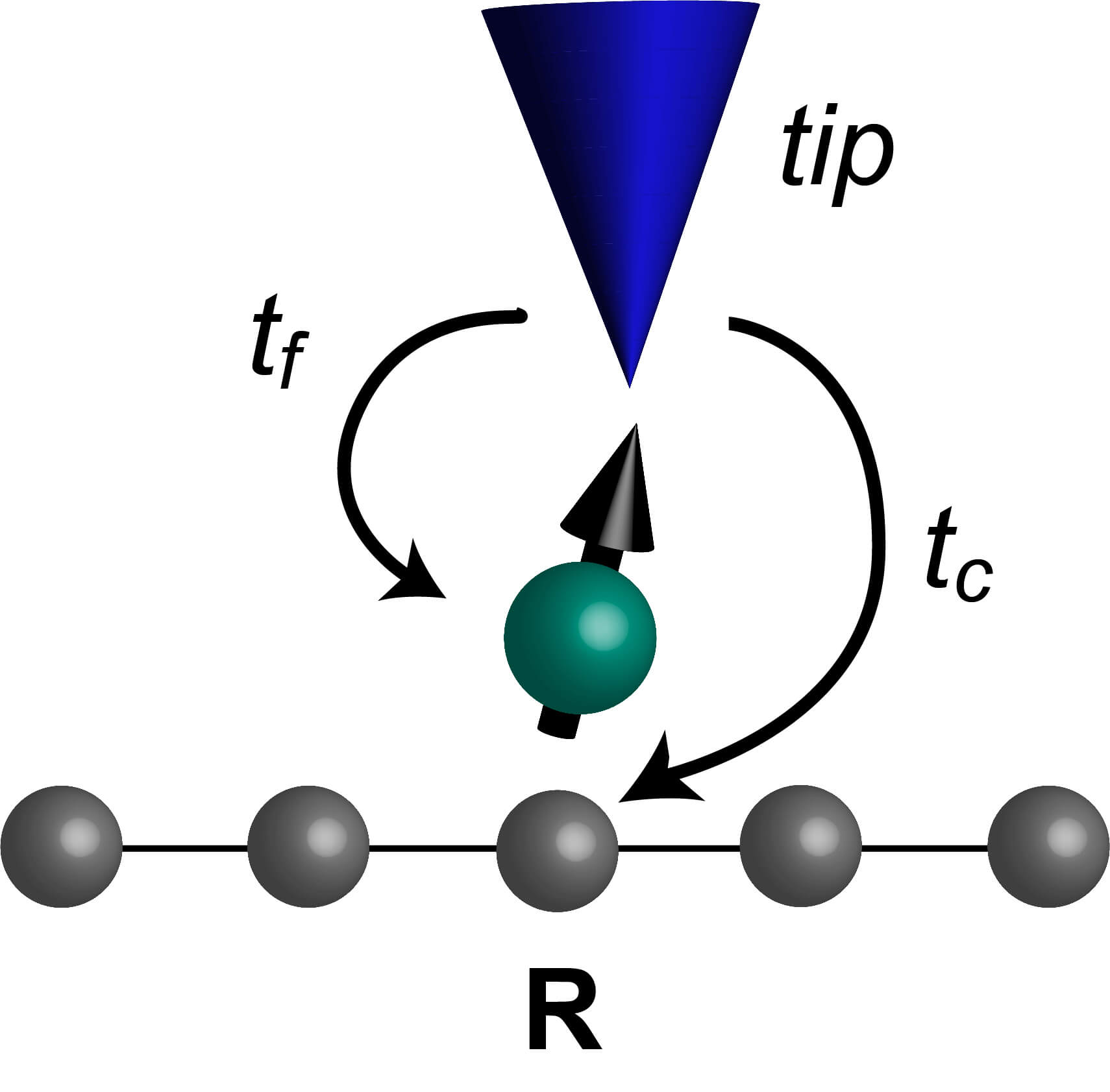}%
 \caption{Paths of electrons tunneling from the STM tip either into the conduction band sites (grey spheres) or into the magnetic level of the Kondo impurity (green sphere), with tunneling amplitudes $t_c$ and $t_f$, respectively.}
 \label{fig:STM}
 \end{figure}

To describe the Kondo screening of the Co adatom by the two-dimensional Au(111) surface state, we employ a large-$N$ expansion \cite{Col83,Hew93,Sen04,Paul07,Read83,Bic87,Aff88}. Here, ${\bf S}^{K}_{\bf R}$ is generalized to $SU(N)$ and represented via
Abrikosov pseudofermions $f^\dagger_{m}, f_{m}$ which obey the
constraint $\sum_{m=1..N} f^\dagger_{m} f_{m}=1$ with $N=2S+1$ being the
spin degeneracy of the magnetic adatom. This constraint is
enforced by means of a Lagrange multiplier $\varepsilon_f$, while
the exchange interaction in Eq.(\ref{eq:H}) is decoupled via the
hybridization field, $s$. The hybridization represents the hopping between the conduction electron states and the pseudofermion $f$-electron states with the resulting Kondo temperature scaling as \cite{Hew93}, $T_K \sim s^2$. For fixed $J$,
$\varepsilon_f$ and $s$ are obtained on the saddle point level by minimizing the
effective action \cite{Read83}. Finally, the tunneling of electrons from the STM tip into the system is described  by the  Hamiltonian
\begin{align}
\hat{H} =  \sum_{\sigma}  t_{c} c^{\dag}_{\textbf{R},\sigma} d_{\sigma} + t_{f} f^{\dag}_{\textbf{R},\sigma} d_{\sigma} + H.c.
\end{align}
where $t_{c}$ ($t_{f}$) are the amplitudes for tunneling of electrons from the tip into the Au(111) surface band (the magnetic $f$-level), as schematically shown in Fig.~\ref{fig:STM}, and $d_{\sigma}$ annihilates a fermion in the STM tip.

To compute the current and associated shot-noise measured by the STM tip, we employ the non-equilibrium Keldysh Greens function formalism \cite{Kel65,Ram86}. Unless otherwise stated, all results presented in Secs.~\ref{sec:SN_Kimp} and \ref{sec:SN_KL} were obtained at zero temperature. When the STM tip is positioned above the magnetic adatom at site ${\bf R}$, the current flowing from the STM tip into the system is given by \cite{Car71}
\begin{eqnarray}
I_{\bf R}(V)&=&-\frac{2e}{\hbar} \, {\rm Re} \, \int_0^{V} \frac{d \omega}{2
\pi} \left[ t_c \, {\hat G}^<_{12}(\omega) + t_f \, {\hat
G}_{13}^<(\omega)  \right] \ ,
 \label{eq:IV}
\end{eqnarray}
with the full lesser Greens function matrix given by
\begin{eqnarray}
{\hat G}^<(\omega) &=& [{\hat 1} - {\hat g}^r(\omega) {\hat t}]^{-1}
{\hat g}^<(\omega) [{\hat 1} -  {\hat t} {\hat g}^a(\omega) ]^{-1} \
; \nonumber \\
{\hat g}^<(\omega) &=& -2i{\hat n_F}(\omega) {\rm Im} \left[ {\hat g}^r(\omega) \right] \ ; \nonumber \\
{\hat g}^r(\omega) &=&
\begin{pmatrix} g^r_t(\omega) & 0 & 0 \\ 0 & g^r_{cc}({\bf R}, {\bf
R}, \omega)& g^r_{cf}({\bf R}, {\bf R},
\omega) \\
0 & g^r_{fc}({\bf R}, {\bf r}, \omega) & g^r_{ff}({\bf R}, {\bf R},
\omega)
\end{pmatrix} \ .
\end{eqnarray}
Here, ${\hat t} $ is the symmetric hopping matrix with non-zero
elements ${\hat t}_{12} = t_c$, ${\hat t}_{13} = t_f$. ${\hat n_F}$
is diagonal containing the Fermi-distribution functions of the tip, $n_F^t(\omega)$, and of the
$f$- and $c$-electron states, $n_F(\omega)$. $g^r_t$ is the retarded Greens
function of the tip, and $g_{\alpha \beta}({\bf r}^\prime, {\bf r},
\tau)=-\langle T_\tau \alpha_{{\bf r}^\prime}(\tau)
\beta^\dagger_{\bf r}(0) \rangle$ ($\alpha,\beta=c,f$) describes the
many-body effects arising from the hybridization of the conduction
band with the $f$-electron level, and the concomitant screening of
the magnetic moment, with
\begin{eqnarray}
g^r_{ff}({\bf R}, {\bf R}, \omega) & = &  \left[\omega-\varepsilon_f
- s^2 g^r_0({\bf R}, {\bf R}, \omega)\right] ^{-1} \ ; \nonumber \\
g^r_{cc}({\bf R}, {\bf R}, \omega) & = & \left\{ \left[ g^r_0({\bf R}, {\bf R},
\omega) \right]^{-1} - \frac{s^2}{\omega-\varepsilon_f + i \delta} \right\}^{-1} \nonumber \\
g^r_{cf}({\bf R}, {\bf R}, \omega) & = & g^r_0({\bf R}, {\bf R},
\omega) s g^r_{ff}({\bf R}, {\bf R}, \omega) \ , \label{eq:GF}
\end{eqnarray}
where $g^r_0$ is the retarded Greens function of the unhybridized
conduction electron band. For a more in-depth discussion, see Ref.~\cite{Morr17}.

It is instructive to consider the weak-tunneling limit ($t_c,t_f \rightarrow 0$) of the current by expanding Eq.(\ref{eq:IV})
up to second order in the tunneling amplitudes, in which case one obtains from Eq.(\ref{eq:IV})
\begin{eqnarray}
I_{\bf R}(V) & =& -\frac{4 \pi e}{\hbar} \pi N_t \int_{-\infty}^{\infty} \frac{d \epsilon}{2\pi} \left[ n^t_F(\epsilon) - n_F(\epsilon) \right] \nonumber \\
& & \hspace{-1.5cm} \times \left[ t_c^2 {\rm Im}g_{cc}^{r}(\epsilon)
 + 2t_c t_f{\rm Im} g_{cf}^{r}(\epsilon) + t_f^2 {\rm Im} g_{ff}^{r}(\epsilon) \right]
\label{eq:I_weak}
\end{eqnarray}
where all $g^r_{\alpha,\beta} \ (\alpha,\beta = c,f)$ are the local retarded Greens' functions at the site of the magnetic adatom, and $N_t$ is the density of states on the tip. We previously demonstrated \cite{Fig10} that the experimental $dI/dV$ lineshape measured at the site of a Co adatom on a Au(111) surface \cite{Mad98} can be described by computing the differential conductance from Eq.(\ref{eq:I_weak}) using the parameters $J=1.39$ eV, $t_f/t_c=-0.066$, and $N=4$. Note that due to a different sign convention for the hybridization $s$ in Ref.\cite{Fig10}, $t_f/t_c$ also changes sign, such that $t_f/t_c=+0.066$ was used in Ref.\cite{Fig10}. These two simultaneous sign changes, however,  do not affect the asymmetry of the $dI/dV$ curves shown below.

We next consider the shot-noise which is defined as the current-current correlation function \cite{Bla00,Zhu03}
\begin{align}
S(t,t') = \langle \{ \delta I(t),\delta I(t') \} \rangle = \langle \{  I(t), I(t') \} \rangle - 2\langle I \rangle ^{2} \ .
\end{align}
We then obtain for the zero frequency noise $S_0 = S(\omega=0)$ at the site of the adatom
\begin{align}
\label{eq:shotnoise}
S_0 & = 2 \bigg(\frac{ie}{\hbar} \bigg)^{2} \int^{\infty}_{-\infty} \frac{d\epsilon}{2\pi} t^{2}_{c} \{ 2 G^{>}_{dc}(\epsilon) G^{<}_{dc}(\epsilon)  -  G^{>}_{dd}(\epsilon) G^{<}_{cc}(\epsilon)  \nonumber \\
& - G^{<}_{dd}(\epsilon) G^{>}_{cc}(\epsilon)\} +2 t_{c} t_{f}  \{G^{>}_{df}(\epsilon) G^{<}_{dc}(\epsilon) +G^{<}_{df}(\epsilon) G^{>}_{dc}(\epsilon) \nonumber \\
&-  G^{>}_{dd}(\epsilon) G^{<}_{fc}(\epsilon) - G^{<}_{dd}(\epsilon) G^{>}_{fc}(\epsilon)\} +  t^{2}_{f} \{2G^{>}_{df}(\epsilon) G^{<}_{df}(\epsilon)  \nonumber \\
& - G^{>}_{dd}(\epsilon) G^{<}_{ff}(\epsilon) - G^{<}_{dd}(\epsilon) G^{>}_{ff}(\epsilon)\}
\end{align}
where $G^{>}$ are the greater Greens functions and all Greens functions in Eq.(\ref{eq:shotnoise}) are local Greens function at the site of the adatom.

Considering again the weak-tunneling limit $t_c, t_f \rightarrow 0$, the expression for the shot noise in Eq.(\ref{eq:shotnoise}) up to second order in the tunneling amplitudes simplifies to
\begin{align}
\label{eq:noise_weak}
S_0 & = -8\pi \bigg(\frac{e}{\hbar} \bigg)^{2} N_{T}\int^{\infty}_{-\infty} \frac{d\epsilon}{2\pi} \nonumber \\
& \left\{ \left[ 1-n_{F}^{T}(\epsilon)\right] n_{F}(\epsilon) + \left[1-n_{F}(\epsilon)\right] n_{F}^{T}(\epsilon) \right\} \nonumber \\
&  \left[ t_c^2 {\rm Im}g_{cc}^{r}(\epsilon)
 + 2 t_c t_f {\rm Im} g_{cf}^{r}(\epsilon) + t_{f}^2 {\rm Im} g_{ff}^{r}(\epsilon) \right] \ .
\end{align}

By comparing the expressions for the current, Eq.(\ref{eq:I_weak}), and shot-noise, Eq.(\ref{eq:noise_weak}) in the weak-tunneling limit, i.e., up to second order in the tunneling amplitudes, we find that at zero temperature the Fano factor, Eq.(\ref{eq:Fano}), is given by $F=1$, implying that the noise is Poissonian. However, the inclusion of higher order tunneling terms in the calculation of the current and shot-noise using Eqs.(\ref{eq:IV}) and (\ref{eq:shotnoise}) respectively, yields not only deviations of $F$ from the Poissonian limit, but also a characteristic bias dependence, that similar to the differential conductance, reflects the Kondo screening process, as shown below.  Finally, we note that the definition of the Fano factor given in Eq.(\ref{eq:Fano}) differs from that used in Refs.\cite{Sela06,Yam11}, as Eq.(\ref{eq:Fano}) involves the total current and noise measured by the STM tip.

To study the shot noise in Kondo lattice systems, we generalize the Hamiltonian in Eq.(\ref{eq:H}) to
\begin{equation}
\hat{H} = - \sum_{\textbf{r},\textbf{r}',\sigma} t_{\textbf{rr}'}c^{\dag}_{\textbf{r},\sigma}c_{\textbf{r}',\sigma}  + J \sum_{\bf r} \textbf{S}_{\textbf{r}}^{K} \cdot \textbf{s}_{\textbf{r}}^{c} + \sum_{\langle {\bf r,r^\prime} \rangle} I_{\bf r, r^\prime} \textbf{S}_{\textbf{r}}^{K} \textbf{S}_{\textbf{r}^\prime}^{K}
\label{eq:H_KL}
\end{equation}
where the sums run over all sites ${\bf r}$ of the conduction electron lattice. The last term represents the antiferromagnetic interaction between the magnetic moments where we assume that $I_{{\bf r,r^\prime}}>0$ is non-zero for nearest-neighbor sites only. Introducing again an Abrikosov pseudo-fermion representation of
${\bf S}^{K}_{\bf r}$, the antiferromagnetic interaction term can be decoupled using $\chi_0=I \langle f^\dagger_{{\bf r},\alpha} f_{{\bf r'},\alpha} \rangle$, which is a measure for the strength of the magnetic correlations in the system.
With this decoupling, the full Green's functions in momentum space, which describe the hybridization between the $c$- and $f$-electron bands, are given by
\begin{eqnarray}
g_{ff}({\bf k},\alpha, \omega) & = &  \left[(g_{ff}^0({\bf k}, \alpha, \omega))^{-1} - s^2
g_{cc}^0({\bf k}, \alpha, \omega) \right] ^{-1} \ ;  \nonumber \\
g_{cc}({\bf k}, \alpha, \omega) & = & \left[(g_{cc}^0({\bf k}, \alpha, \omega))^{-1} - s^2
g_{ff}^0({\bf k}, \alpha, \omega) \right] ^{-1} \ ;  \nonumber \\
g_{cf}({\bf k}, \alpha, \omega) & = & - g_{cc}^0({\bf k}, \alpha, \omega) s
g_{ff}({\bf k}, \alpha, \omega) \ , \label{eq:GF_KL}
\end{eqnarray}
where
\begin{align}
g_{cc}^0 & = \frac{1}{\omega + i \delta -\varepsilon^c_{\bf k}} \nonumber \\
g_{ff}^0 &= \frac{1}{\omega + i \delta -\varepsilon^f_{\bf k}} \nonumber \\
\varepsilon^f_{\bf k} & =-2 \chi_0 (\cos{k_x} + \cos{k_y} ) + \varepsilon_f \nonumber \\
\varepsilon^c_{\bf k} & =-2 t (\cos{k_x} + \cos{k_y} ) - \mu_c \ .
\end{align}
Here, $\varepsilon^f_{\bf k}$ and $\varepsilon^c_{\bf k}$ are the dispersions of the unhybridized conduction electron and $f$-electron bands,  respectively. The dispersions of the hybridized conduction and $f$-electron bands are then given by
\begin{align}
\label{eq:EKpm}
E_{\bf k}^\pm = \frac{\varepsilon^c_{\bf k}+\varepsilon^f_{\bf k}}{2} \pm \sqrt{ \left( \frac{\varepsilon^c_{\bf k}-\varepsilon^f_{\bf k}}{2} \right)^2 + s^2} \ .
\end{align}
Finally, we note that the formal expressions for the current and the shot-noise in the Kondo lattice are the same as given in Eqs.(\ref{eq:IV}) and (\ref{eq:shotnoise}), respectively, with the local Greens functions in Eq.(\ref{eq:GF}) being computed via Fourier transform from their momentum space form in Eq.(\ref{eq:GF_KL}).

\section{Shot Noise around a Kondo impurity}
\label{sec:SN_Kimp}

We begin by considering the current and shot-noise around a single Kondo impurity, using the parameters previously employed to explain the differential conductance of a Kondo-screened Co adatom located on a Au(111) surface \cite{Mad98,Fig10}. While the Fano factor is unity in the weak tunneling limit, i.e., up to second order in the tunneling amplitudes, it deviates from this result with increasing tunneling amplitude $t_c, t_f$, exhibiting a characteristic lineshape that sensitively depends on the ratio of the tunneling amplitudes $t_f/t_c$.  Thus, in order to be able to measure experimentally a characteristic Fano factor, it is desirable to have large tunneling amplitudes, corresponding to small distances between STM tip and sample, and hence sufficiently large currents.
\begin{figure}[h]
\includegraphics[width=8.5cm]{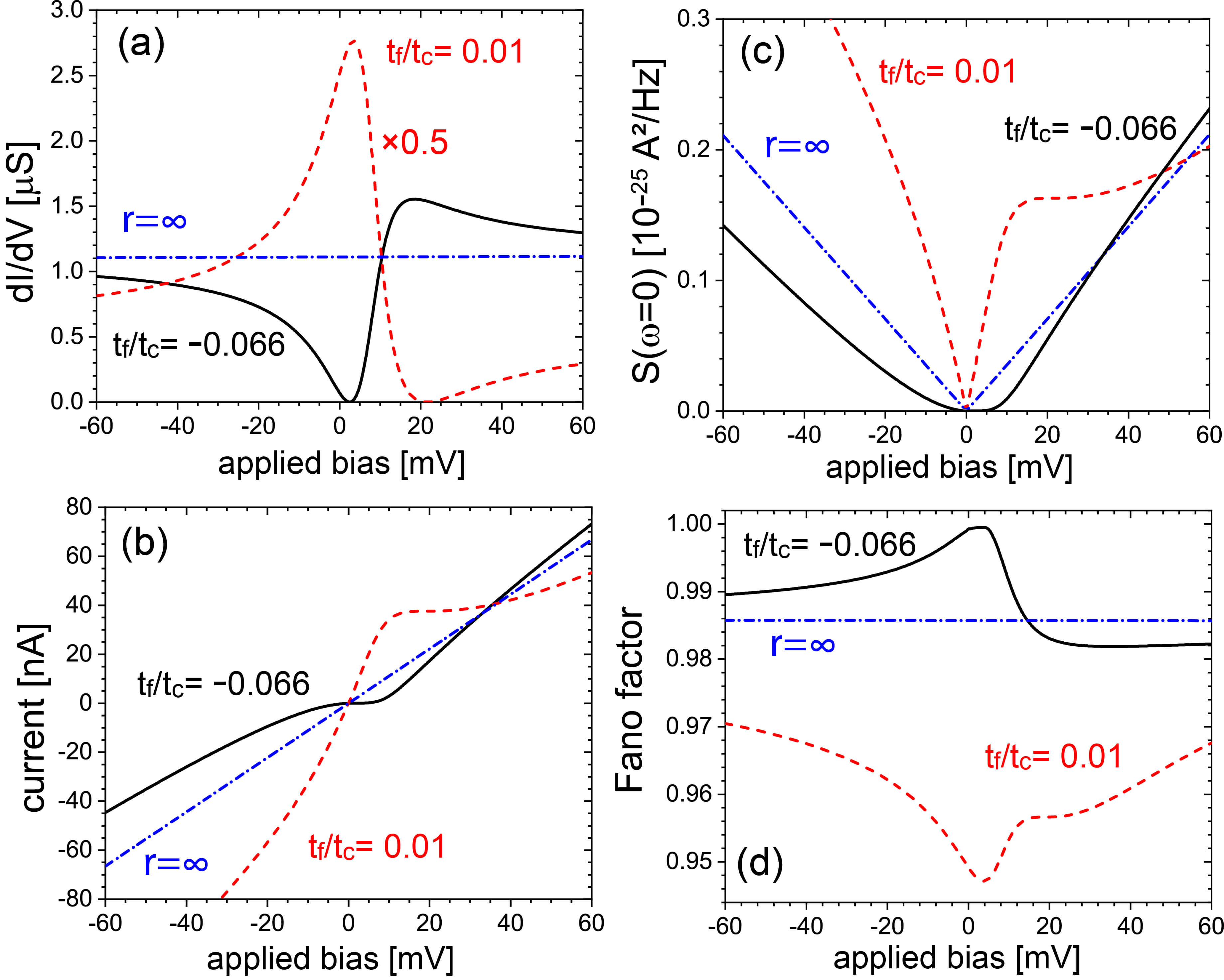}%
 \caption{(a) $dI/dV$,  (b) current $I$, (c) noise $S_0$, and (d) Fano factor $F$ for two different values of $t_f/t_c=-0.066$ and $t_f/t_c=0.01$, as well as away from the Kondo impurity at $r = \infty$. Results are shown for zero temperature. }
 \label{fig:Kimp_comp}
 \end{figure}
While the ratio $t_f/t_c$ can be determined by fitting the experimental $dI/dV$ lineshape, as was done for the case of a Co adatom on a Au(111) surface in Ref.~\cite{Fig10}, it is difficult to extract the absolute values of the tunneling amplitudes. Therefore, in order to determine whether deviations of $F$ from unity can be observed experimentally, it is necessary to treat $t_c, t_f$ as implicit parameters, and correlate the $IV$-curves that result from given values for $t_c, t_f$ with the form of the Fano factor. We therefore present below the current, differential conductance and noise in absolute units for different sets of $t_c, t_f$. We note that increasing $t_c$ with constant $t_f/t_c$ leads to an increase in the current between the tip and the system, and thus corresponds to decreasing the distance between the STM tip and the sample in experiments. Current state-of-the-art STS experiments can achieve currents in the tunneling regime of hundreds of nA for a bias of a few mV \cite{Kim15,Kim17}, rendering all theoretical results shown below within the experimental accessible region.

\begin{figure}[h]
\includegraphics[width=8.5cm]{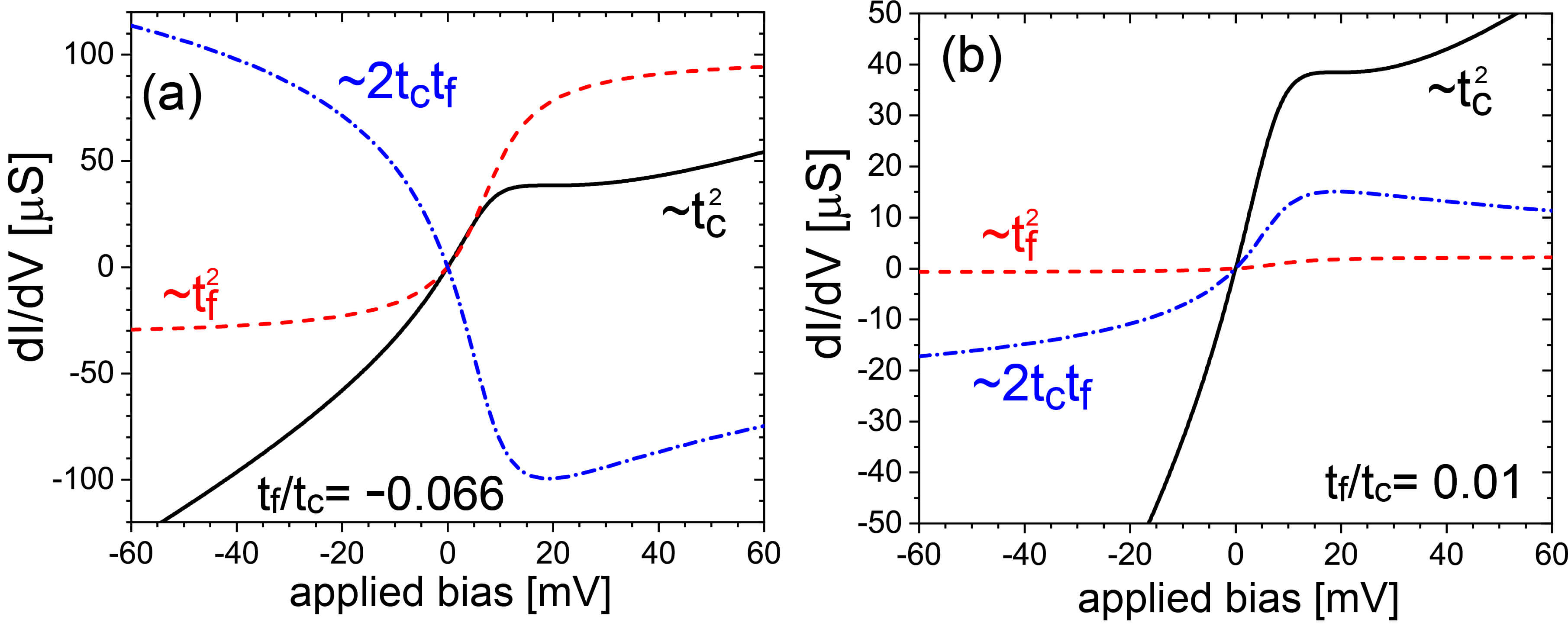}%
 \caption{The three contributions to the total current from the weak-tunneling limit of Eq.(\ref{eq:I_weak}) which are proportional to $t_c^2,t_f^2$ and $2t_ct_f$ for (a) $t_f/t_c=-0.066$, and (b) $t_f/t_c=+0.01$.}
 \label{fig:Kimp_3I}
 \end{figure}

Using the same set of parameters as previously employed in Ref.\cite{Fig10}, we present in Fig.\ref{fig:Kimp_comp}(a) the differential conductance at the site of a single Kondo impurity for $t_c=0.1$eV and two values of $t_f/t_c$. For $t_f/t_c=-0.066$, we obtain the $dI/dV$ lineshape (black line) that was previously employed to fit the experimental lineshape measured above a Co adatom on a Au(111) surface. For comparison, we also present (i) $dI/dV$ for $t_f/t_c=+0.01$ (red dashed line), whose lineshape exhibits an asymmetry that is reversed from that obtained for $t_f/t_c=-0.066$, and (ii) $dI/dV$ away from the adatom at $r=\infty$ (blue dotted-dashed line) which is that of the unhybridized conduction band. To understand the difference in the asymmetry of the $dI/dV$ lineshapes, we consider the $IV$-curves for these three cases in Fig.\ref{fig:Kimp_comp}(b). We find that the Kondo correlations lead to a suppression of the current for $t_f/t_c=-0.066$ from its value at $r=\infty$, but to an enhancement for $t_f/t_c=+0.01$. This in turn accounts for the change in the asymmetry of the differential conductance curves between $t_f/t_c=-0.066$ and $+0.01$. The origin of this enhancement/suppression can be understood from the weak tunneling limit of the current, Eq.(\ref{eq:I_weak}), as it lies in the interference term $\sim t_c t_f$. For $t_f<0$, this interference term leads to a backflow of current from the system into the tip, reducing the overall magnitude of the current, as shown in Fig.~\ref{fig:Kimp_3I}(a) where we present the contributions to the total current arising from the three terms proportional to $t_c^2,t_f^2$ and $2t_ct_f$ in Eq.(\ref{eq:I_weak}). In contrast, for $t_f>0$, the interference term leads to an additional current flowing from the tip into the system  as shown in Fig.~\ref{fig:Kimp_3I}(b), thus increasing the total current.
It also follows from the $IV$-curves that the magnitude of the current for bias of a few mV falls within the experimentally accessible range, implying that the value of $t_c=0.1$eV is experimentally achievable.

In Fig.\ref{fig:Kimp_comp}(c) we present the zero-frequency shot-noise, $S_0$, for the cases $t_f/t_c=-0.066,+0.01$ and $r=\infty$. Similar to the current, we find that the Kondo correlations either suppress (for $t_f/t_c=-0.066$) or enhance (for $t_f/t_c=+0.01$) the shot-noise with respect to its form at $r=\infty$. The reason for this suppression or enhancement is similar to that for the current: due to the backflow of the current from the system into the tip arising from the interference term for $t_f/t_c<0$, the contribution to the noise arising from the current-current correlation between the current flowing directly from the tip into the system, and the backflow is negative, thus reducing the overall noise. In contrast, for $t_f/t_c>0$ the contribution to the noise $\sim t_c t_f$ is positive, leading to an enhanced noise in the vicinity of the Kondo resonance.

Finally, in Fig.\ref{fig:Kimp_comp}(d) we present the Fano factor for all three cases, which exhibits a peak near the Kondo resonance for $t_f/t_c=-0.066$, and a dip for $t_f/t_c=0.01$. Moreover, the Fano factor for $t_f/t_c=-0.066$ near the Kondo resonance is enhanced over its value for $r=\infty$, while it is suppressed for $t_f/t_c=0.01$. To understand this difference in the Fano factor near $V=0$, we consider the Landauer formula \cite{Lan57} for the current
\begin{equation}
I=\frac{e^2}{\pi \hbar} V T_{eff}
\label{eq:I_Lan}
\end{equation}
where $T_{eff}$ is the effective transmission coefficient between the tip and the system. A comparison with the weak-tunneling expression for the current, Eq.(\ref{eq:I_weak}), shows that to leading order in $V$
\begin{align}
T_{eff} &= -2\pi N_t \left[ t_c^2 {\rm Im}g_{cc}^{r}(\varepsilon_F)
 + 2t_c t_f{\rm Im} g_{cf}^{r}(\varepsilon_F) \nonumber \right. \\
 & \hspace{1cm}  \left. + t_f^2 {\rm Im} g_{ff}^{r}(\varepsilon_F) \right] \nonumber \\
 & = \frac{\pi \hbar }{e^2} \left. \frac{dI}{dV} \right|_{V=0}
 \label{eq:T_eff}
\end{align}
It follows from Fig.~\ref{fig:Kimp_comp}(a) that $T_{eff} \sim \left. \frac{dI}{dV} \right|_{V=0}$  is smaller for $t_f/t_c=-0.066$ than for $t_f/t_c=0.01$. Similarly, the shot-noise can be written in terms of $T_{eff}$ as \cite{Khl87}
\begin{equation}
S_0=\frac{2e^3}{\pi \hbar} |V| T_{eff}(1-T_{eff}) \ .
\label{eq:S_Teff}
\end{equation}
A comparison of Eq.(\ref{eq:S_Teff}) with the weak tunneling limit for $S_0$ in Eq.(\ref{eq:noise_weak}) yields the same $T_{eff}$ as in Eq.(\ref{eq:T_eff}) to leading order in $t_c,t_f$. We note that the term $\sim T_{eff}^2$ in Eq.(\ref{eq:S_Teff}) scales as the hopping amplitudes to the fourth power, and is therefore not contained in the weak-tunneling limit of $S_0$ in Eq(\ref{eq:noise_weak}). By combining Eqs.(\ref{eq:I_Lan}) and (\ref{eq:S_Teff}), we obtain for the Fano factor near $V=0$ $F=(1-T_{eff})$, which is thus larger for $t_f/t_c=-0.066$ than for $t_f/t_c=0.01$, in agreement with our numerical results shown in Fig.\ref{fig:Kimp_comp}(d). We thus conclude that there exist an interesting correlation between the lineshape of the Kondo resonance (as determined by $t_f/t_c$) and the enhancement or suppression of the Fano factor with respect to the $r=\infty$ result.

\begin{figure}[h]
\includegraphics[width=8.5cm]{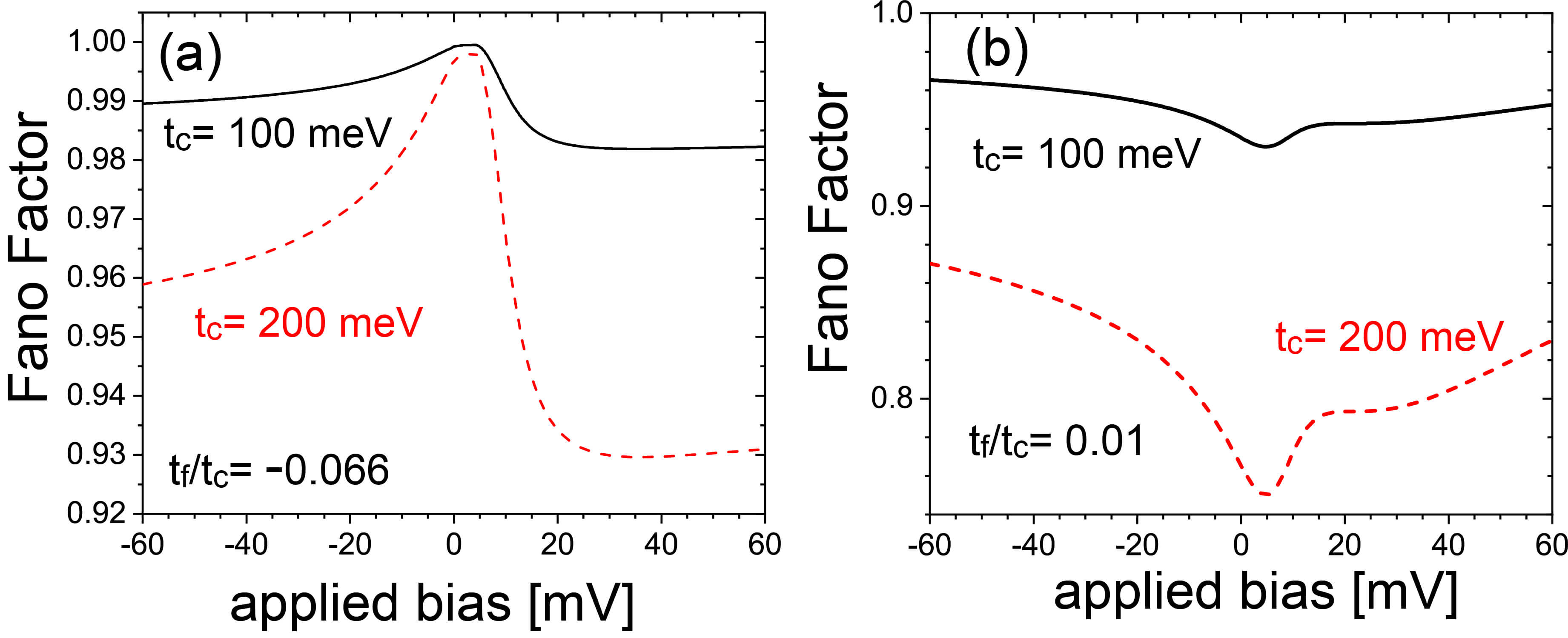}%
\caption{Evolution of the Fano factor $F$ with increasing $t_c$ for (a) $t_f/t_c=-0.066$, and (b) $t_f/t_c=0.01$}
\label{fig:evol}
\end{figure}
A unique feature of the Fano factor is that its overall lineshape, i.e, its bias dependence, is essentially independent of $t_c$, varying only with $t_f/t_c$. To demonstrate this, we present in Fig.~\ref{fig:evol} the Fano factor $F$ for several values of $t_c$ with constant $t_f/t_c$.
While the overall lineshape of the Fano factor does not change with increasing $t_c$ (for constant $t_f/t_c$), its overall variation increases, thus becoming easier to observe experimentally. It is interesting to note that the maximum of the Fano factor for $t_f/t_c=-0.066$ remains close to unity near the Kondo resonance, implying that the transmission amplitude $T_{eff}$ remains approximately zero. On the other hand, for $t_f/t_c=+0.01$, the suppression of the Fano factor near the Kondo resonance increases, implying that $T_{eff}$ increases with increasing $t_c$.

\begin{figure}[h]
\includegraphics[width=8.5cm]{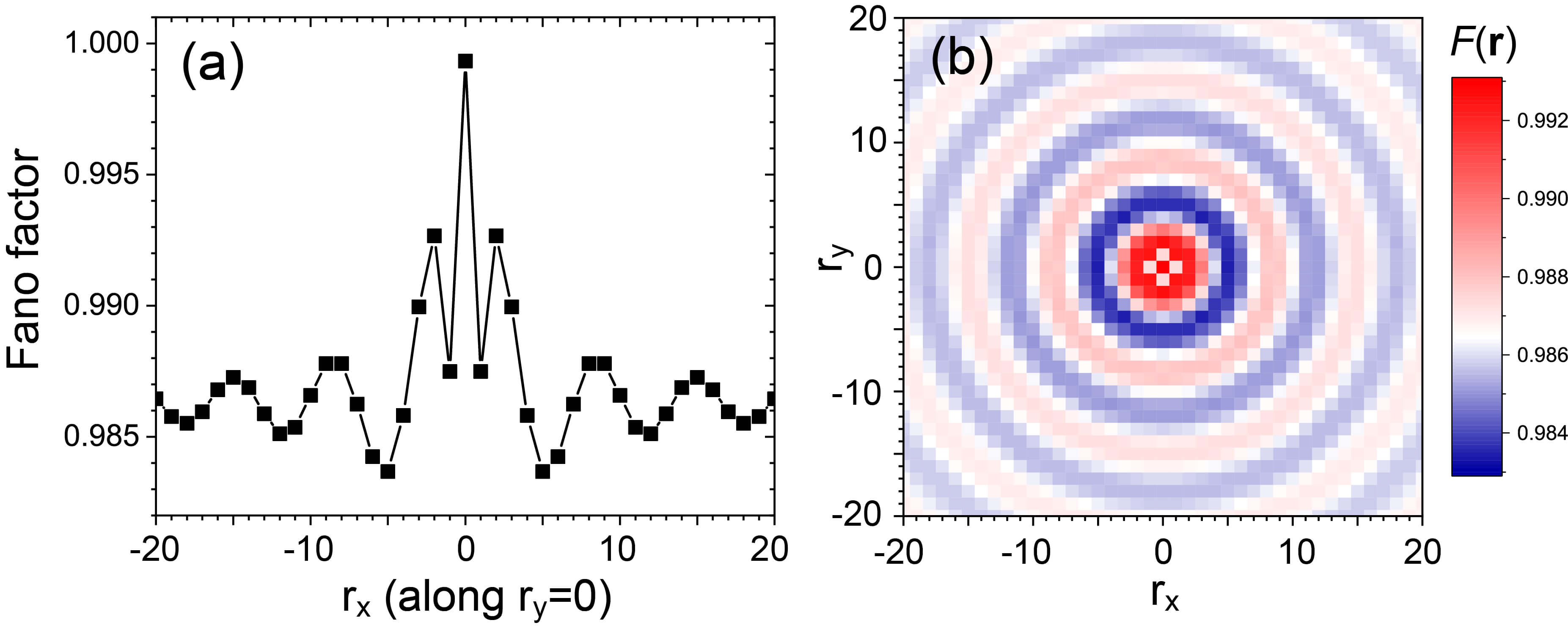}%
 \caption{(a) Linecut of $F$ through the site the magnetic adatom for $t_c=0.1$ eV, $V=5$mV and $t_f/t_c=-0.066$. (b) Spatial plot of $F$.}
 \label{fig:Fano_spatial}
 \end{figure}
The Fano factor exhibits spatial oscillations, as shown in Fig.~\ref{fig:Fano_spatial}, where we present a linecut of the Fano factor through the magnetic adatom [Fig.~\ref{fig:Fano_spatial}(a)] as well as a spatial plot of $F({\bf r})$ [Fig.~\ref{fig:Fano_spatial}(b)].
The spatial plot of $F({\bf r})$ reveals nearly isotropic oscillations whose wavelength is given by $\lambda \approx 6.5 a_0$ which is half of the Fermi wave-length. We can therefore conclude that the spatial oscillations of the Fano factor are $2k_F r$-oscillations, arising from scattering of the surface conduction electrons from the magnetic adatom. Similar spatial oscillations in the conductance fluctuations were interpreted as a signature of the Kondo screening cloud \cite{Pat09}.

\begin{figure}[h]
\includegraphics[width=8.5cm]{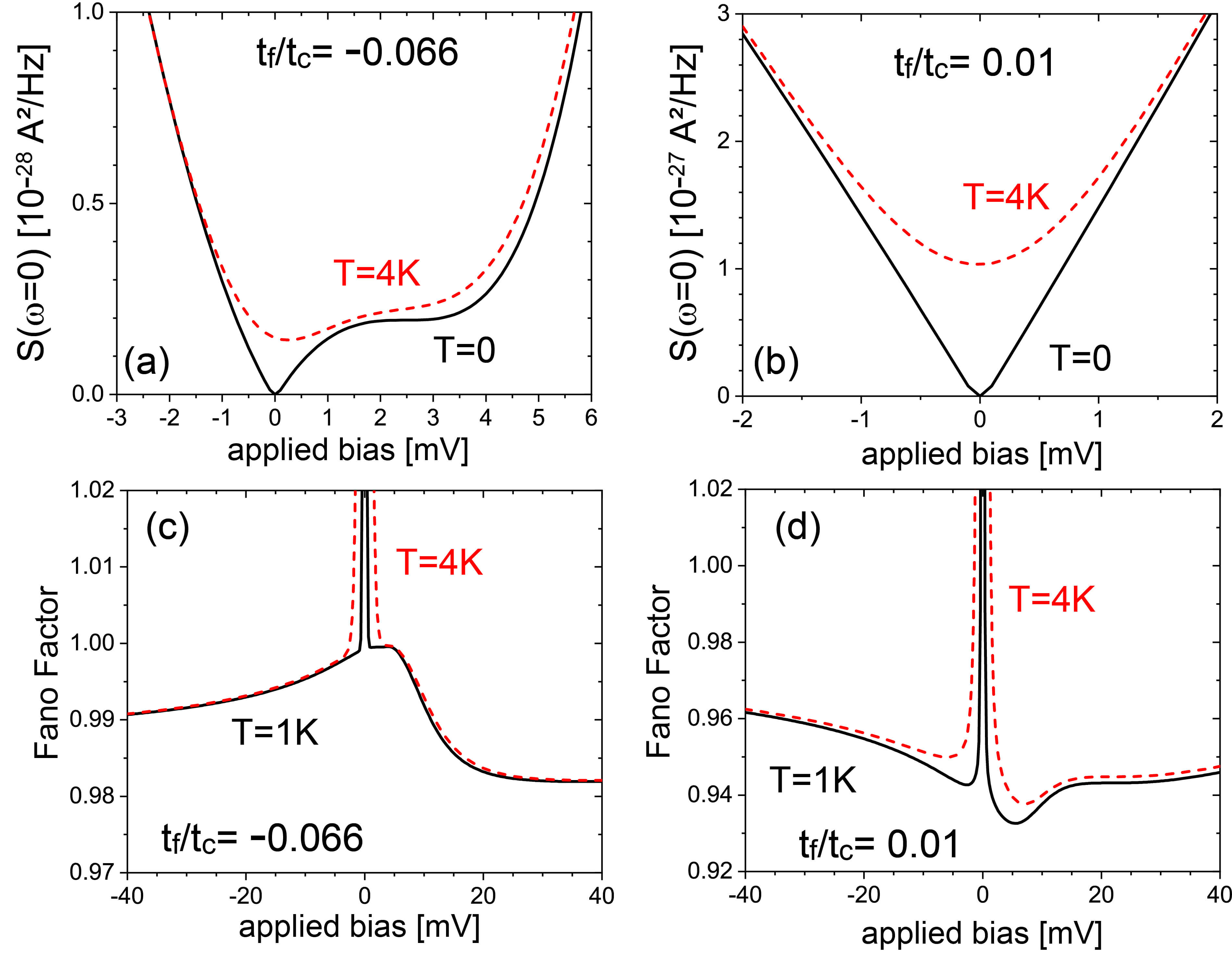}%
 \caption{Comparison of the zero-frequency noise, $S(\omega=0)$, at $T=0$ and $T=4K$ for $t_c=0.1$eV and (a) $t_f/tc=-0.066$, and (b) $t_f/tc=0.01$ (note the different $x$- and $y$-axes scales). Temperature evolution of the Fano factor for $t_c=0.1$eV and (c) $t_f/tc=-0.066$, and (d) $t_f/tc=0.01$.  }
 \label{fig:Fano_temp}
 \end{figure}
Finally, we briefly comment on the temperature dependence of the Fano factor. For any non-zero temperature, there are thermal contributions to the zero-frequency noise which are non-zero even at $V=0$, as shown in Figs.~\ref{fig:Fano_temp}(a) and (b) for $t_f/tc=-0.066$ and $0.01$, respectively (note the different $x$- and $y$-axes scales). On the other hand, the current vanishes for $V=0$, independent of temperature. This implies that for any non-zero temperature, the Fano factor exhibits a divergence at $V=0$, as shown in Fig.~\ref{fig:Fano_temp}(c) and (d). We note that the bias range over which the Fano factor at $T=4K$ is enhanced over its $T=0$ value varies significantly with $t_f/t_c$.

\section{Shot noise in a Kondo lattice}
\label{sec:SN_KL}

\begin{figure}[h]
\includegraphics[width=8cm]{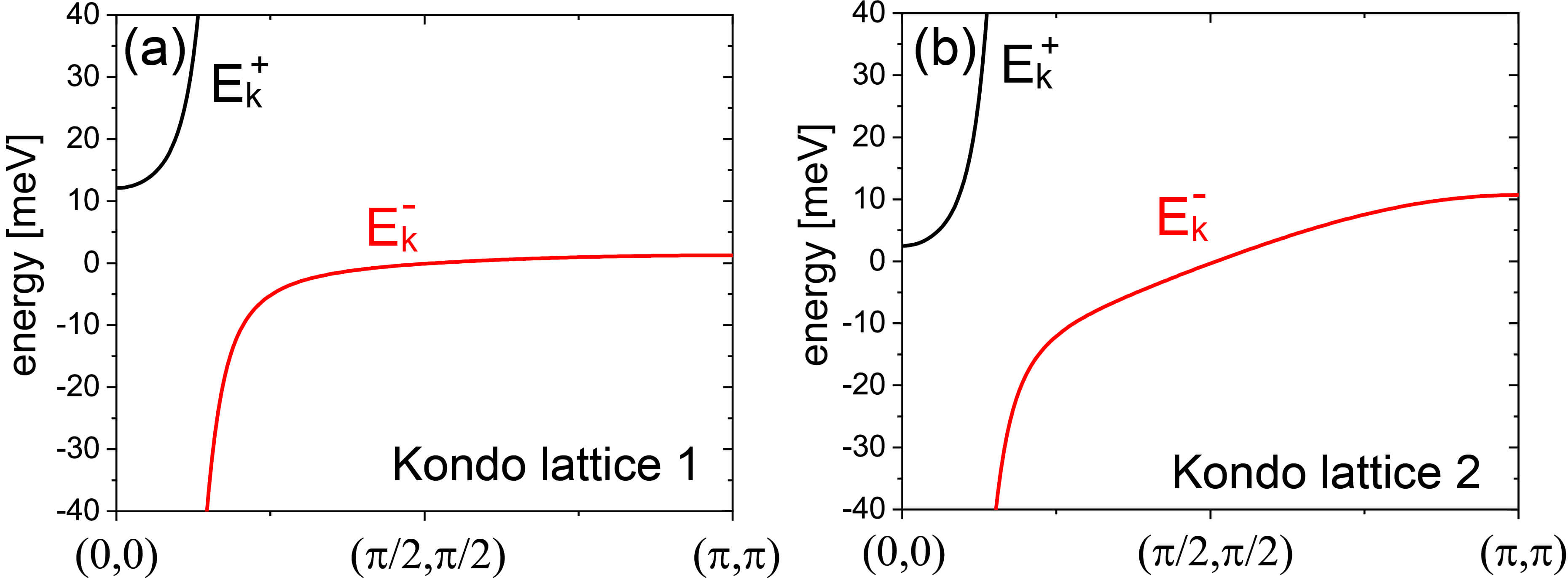}%
 \caption{The dispersions $E_{\bf k}^{\pm}$ from Eq.(\ref{eq:EKpm}) along $(0,0) \rightarrow (\pi,\pi)$ with $t=500meV$, $\mu=-3.618t$, $N = 2$, $J = 500$meV, $N_t = 1eV^{-1}$, for (a) Kondo lattice 1 with $I/J$ = 0.001 yielding $s = 48.5$meV, $\varepsilon_f = 1.2$meV,  and $\chi_0 =0.17$meV, and (b) Kondo lattice 2 with $I/J$ = 0.015 yielding $s = 48.0$meV, $\varepsilon_f = 0.94$meV, and $\chi_0 =2.59$meV.  }
 \label{fig:KL_dispersion}
 \end{figure}

\begin{figure}[h]
\includegraphics[width=8cm]{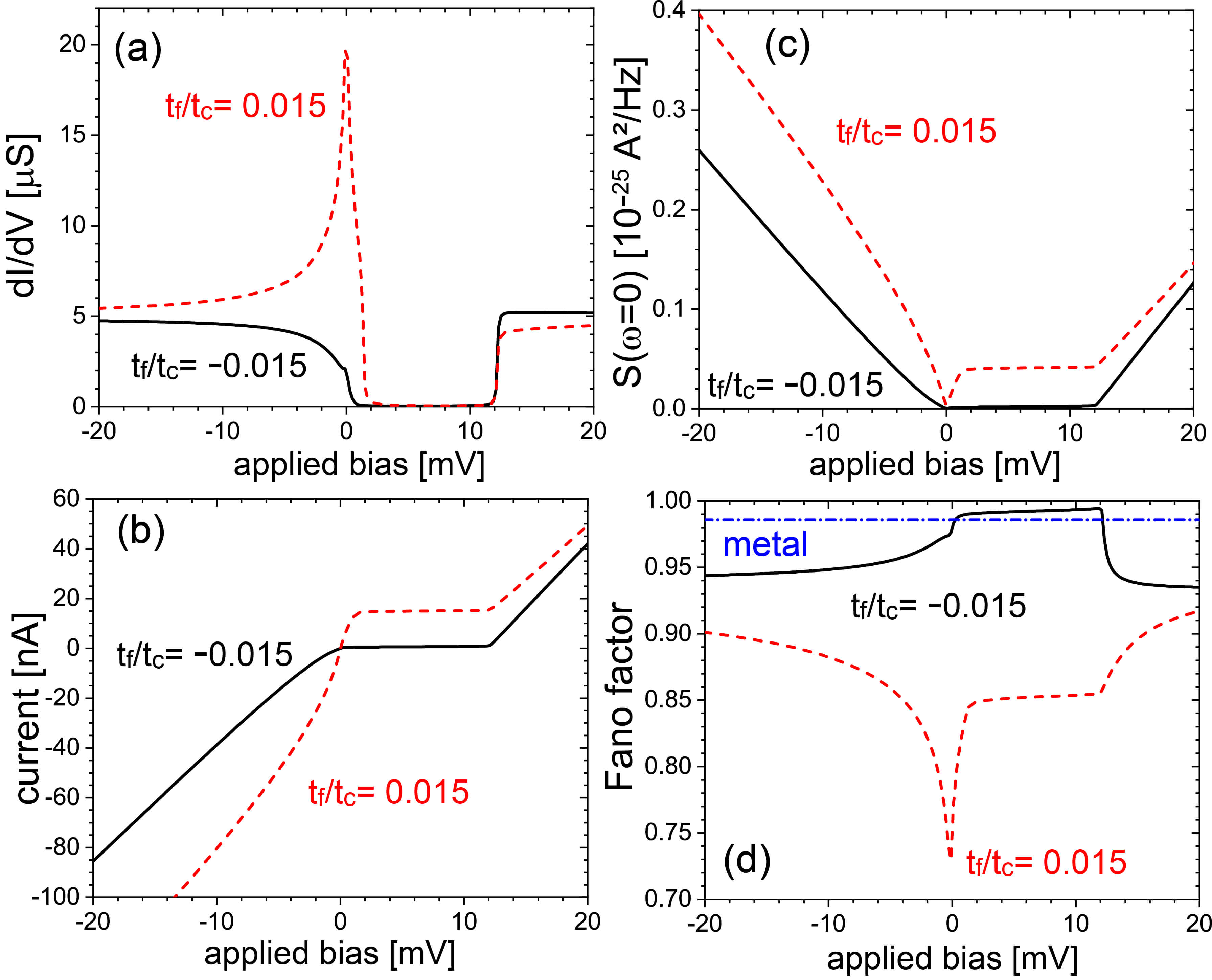}%
 \caption{For Kondo lattice 1: (a) $dI/dV$, (b) current, (c) noise, and (d) Fano factor with $t_c=0.1$ eV and two different values of $t_f/t_c$.}
 \label{fig:KL1}
 \end{figure}

We next study the form of the current and shot-noise in a Kondo lattice. To this end, we consider two different sets of parameters for the Kondo lattice model of Eq.(\ref{eq:H_KL}) previously considered in Ref.\cite{Fig10}: one in which the antiferromagnetic interaction is sufficiently small [$I/J=0.001$, Kondo lattice 1 (KL1)], such that the system exhibits a hard hybridization gap [see Figs.~\ref{fig:KL_dispersion}(a) and \ref{fig:KL1}(a)], and one in which the antiferromagnetic interaction is strong enough [$I/J=0.015$, Kondo lattice 2 (KL2)] such that the system's dispersion does not any longer exhibit an indirect gap [see Fig.~\ref{fig:KL_dispersion}(b)] and the hybridization gap is seen as a suppression in $dI/dV$ rather than hard gap (see Fig.~\ref{fig:KL2}(a), for a more in-depth review, see Ref.~\cite{Morr17}).

We begin by considering the form of the noise and Fano factor for Kondo lattice 1 and present in Fig.~\ref{fig:KL1}(a) the differential conductance for two different values of $t_f/t_c= \pm 0.015$. As expected, $dI/dV$ exhibits a hard hybridization gap, and very different asymmetries for the two values of $t_f/t_c$, similar to the case of a single Kondo impurity shown in Fig.~\ref{fig:Kimp_comp}.
In Figs.~\ref{fig:KL1}(b) and (c), we present the resulting current and shot-noise. Both the current and the shot-noise are bias independent inside the hybridization gap, but overall show a very similar bias dependence to that of the single Kondo impurity. Finally, in Fig.~\ref{fig:KL1}(d) we show the resulting Fano factor. Similar to the single Kondo impurity, the Fano factor is correlated with the asymmetry of the differential conductance. For $t_f/t_c=-0.015$, the Fano factor is close to unity in the hybridization gap, implying that the transmission coefficient is small. In contrast, for $t_f/t_c=+0.015$, the Fano factor is strongly suppressed near the hybridization gap, implying a much larger transmission coefficient. Comparing the Fano factor with that of an uncorrelated metal shows that the strong correlations arising from Kondo screening lead to an overall suppression of the Fano factor independent of the value of $t_f/t_c$, except for the immediate vicinity of the hybridization gap for $t_f/t_c=-0.015$, where the Fano factor is slightly larger than that of the metallic systems.

\begin{figure}[h]
\includegraphics[width=8cm]{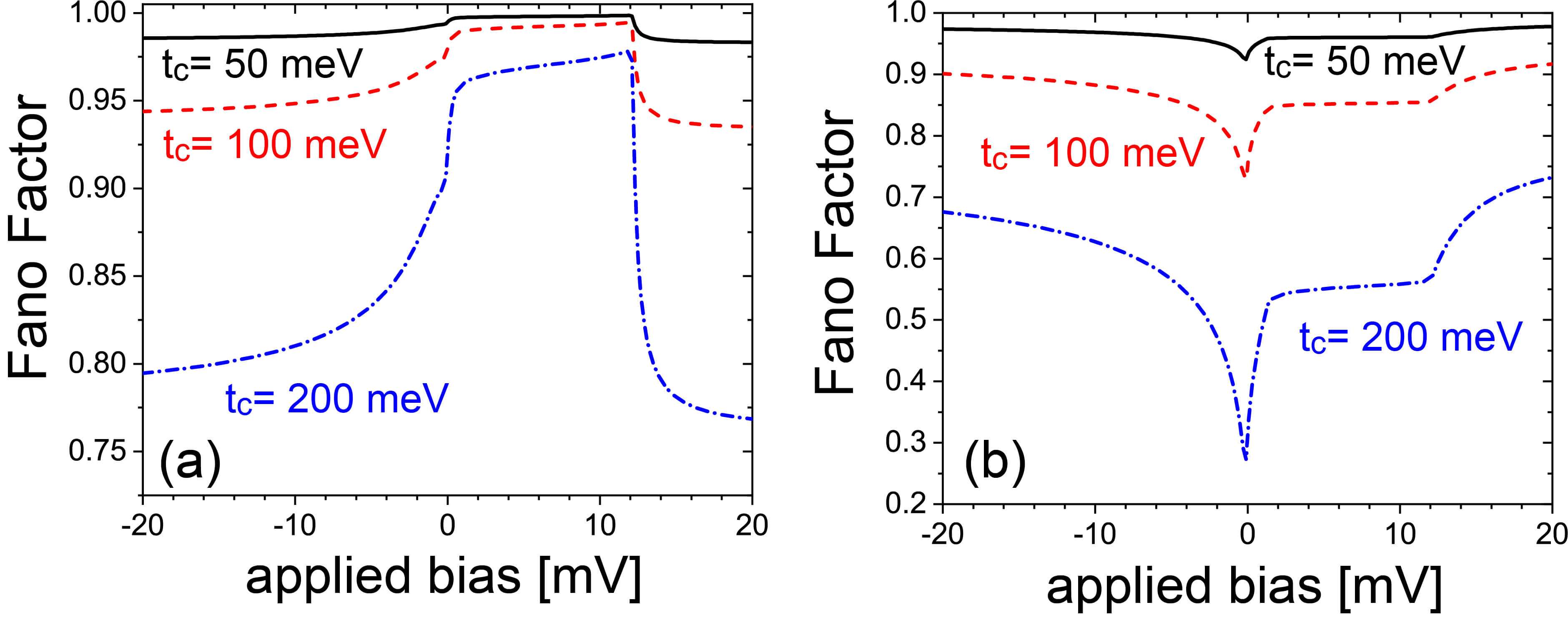}%
\caption{Evolution of the Fano factor $F$ with increasing $t_c$ in Kondo lattice 1 for (a) $t_f/t_c=-0.015$, and (b) $t_f/t_c=0.015$.}
\label{fig:KL_evolution}
\end{figure}

\begin{figure}[h]
\includegraphics[width=8cm]{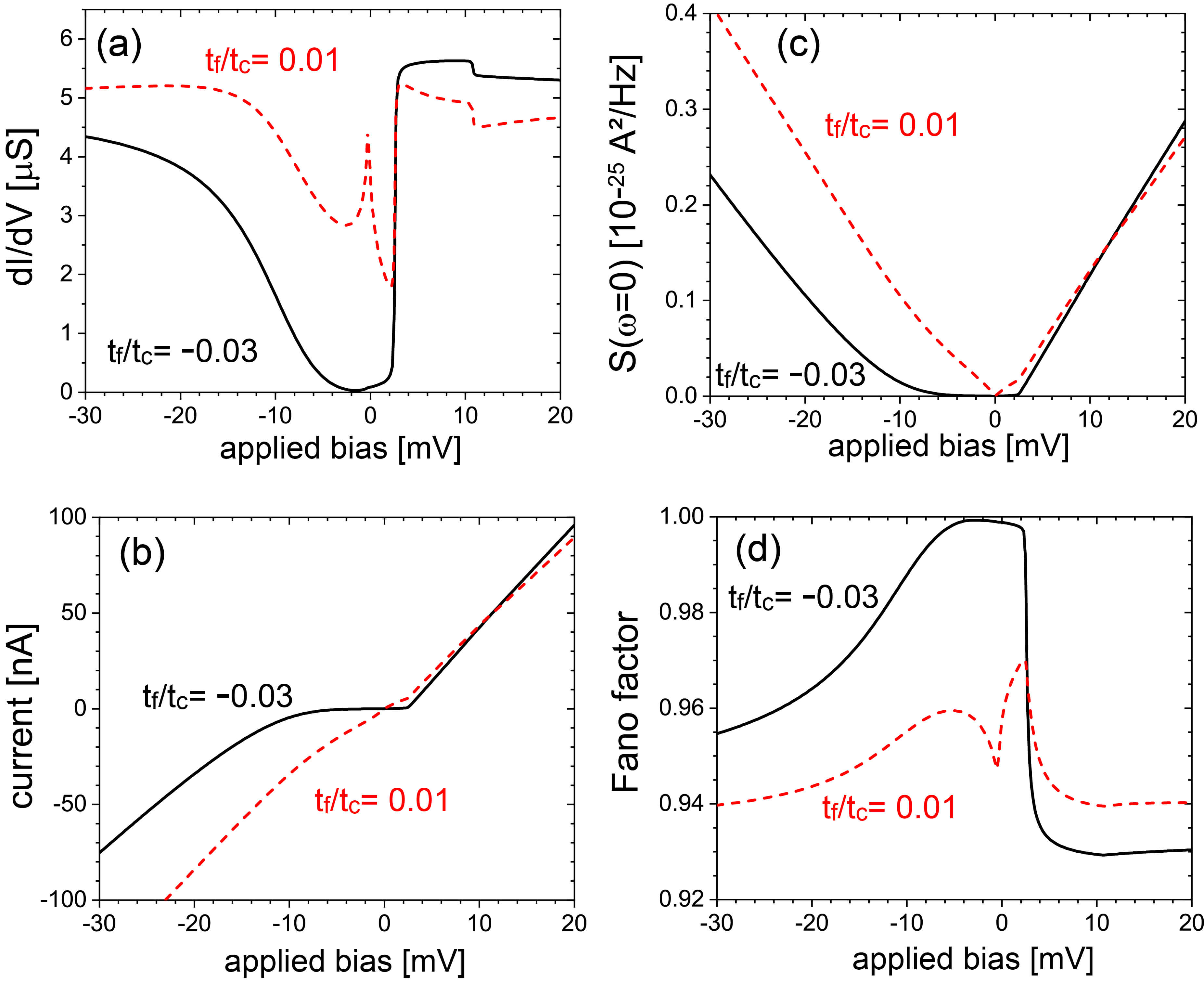}%
 \caption{For Kondo lattice 2: (a) $dI/dV$, (b) current, (c) noise, and (d) Fano factor with $t_c=0.1$ eV and two different values of $t_f/t_c$.}
 \label{fig:KL2}
 \end{figure}

Similar to the case of the single impurity, we find that the overall shape of the Fano factor is independent of the tunneling amplitudes $t_c,t_f$ (for fixed $t_f/t_c$), as shown in Fig.~\ref{fig:KL_evolution}, and that only the overall variation of the Fano factor increases with increasing tunneling amplitudes.

We next consider the form of the noise and Fano factor in Kondo lattice 2, and present in Fig.~\ref{fig:KL2}(a) the resulting differential conductance for two different values of $t_f/t_c=-0.03,0.01$. The larger antiferromagnetic interaction (in comparison to KL1), and the resulting larger value of $\chi_0$, give rise to two interesting effects: (a) $dI/dV$ does not any longer show a hard hybridization gap, but only a suppression, and (b) the van-Hove singularity of the heavy $f$-electron band has been moved inside the hybridization gap, as particularly evident for $t_f/t_c=0.01$. Both features are qualitatively similar to the ones found in the differential conductance of the heavy fermion material URu$_2$Si$_2$ \cite{Ayn10,Yuan12}.
Similar to the $dI/dV$, the current and shot-noise shown in Figs.~\ref{fig:KL2}(b) and (c) differ significantly for negative bias $V<0$, while being quite similar for positive bias $V>0$. In Fig.~\ref{fig:KL2}(d) we present the resulting Fano factor. For $t_f/t_c=-0.03$ where the suppression of $dI/dV$ is more pronounced, the Fano factor is close to unity, implying a vanishing $T_{eff}$. Interestingly enough, the van-Hove singularity inside the hybridization gap leads to a strong suppression of the Fano factor for $t_f/t_c=0.015$. This strong correlation between the form of the differential conductance and the Fano factor represents an important test for future STSNS experiments.\\

\section{Conclusions}
\label{sec:concl}

In conclusion, we have investigated the relation between the differential conductance, current, shot-noise and the resulting Fano factor measured via shot noise scanning tunneling spectroscopy around a single Kondo impurity as well as in Kondo lattices. We demonstrated that Kondo screening leads to a characteristic lineshape of the Fano factor, that is similar to the Kondo resonance observed in the differential conductance. Moreover, the lineshape of $F$  is strongly dependent on the ratio of the tunneling amplitudes $t_f/t_c$ and can be enhanced or suppressed due to interference effects arising from tunneling into the conduction and $f$-electron levels. As such, it is not only a sensitive probe for the correlation effects arising from Kondo screening, but also for quantum interference between tunneling electrons. Moreover, we showed that near the Fermi energy, there exists a correlation between the form of $dI/dV$ and $F$ through the effective transmission coefficient $T_{eff}$, such that a suppression in $dI/dV$ leads to a value of $F$ near unity, while a peak in $dI/dV$ gives rise to a strong suppression in $F$. We also predicted that around a single Kondo impurity, the Fano factor exhibits spatial oscillations whose wavelength arises from $2k_Fr$ oscillations of the scattered conduction electrons. In Kondo lattices, we find that the Fano factor possesses a correlation with the differential conductance that is similar to that in the single Kondo impurity case. This correlation represents a prediction of the effects of quantum interference arising from multiple tunneling paths that could be tested in future STSNS experiments.

\acknowledgements

We would like to thank M. Allan, R. Berndt, H. Kim and F. Massee for stimulating discussion. This work was supported by the U. S. Department of Energy, Office of Science, Basic Energy Sciences, under Award No. DE-FG02-05ER46225.

\end{document}